# The Bias Paradox: How AI Personas Can Overcome Human Limitations in UX Research

Ozgur Taylan Celik
Fibabanka R&D Center
Istanbul Turkey
ozgurtaylan.celik@fibabanka.com.tr

## ABSTRACT

This position paper examines a paradox encountered in UX research practice: a situation where real human participants delivered less authentic insights than AI personas might have, due to context-induced biases. We share our experience developing research-based AI personas using ChatGPT's CustomGPT and conducting a design thinking workshop with high-net-worth banking clients. The workshop setting, including a luxury hotel, present portfolio managers, and hospitality dynamics, introduced biases that compromised the feedback. We propose that AI personas offer an underexplored opportunity to mitigate certain human limitations in user research, and call for frameworks that help teams recognize when traditional research contexts introduce biases that AI personas might help avoid.

## 1 Introduction

Discussions of AI personas in human-centered design typically emphasize efficiency and scale [5, 7]. Traditional personas often remain static artifacts that teams struggle to integrate into ongoing design decisions [2, 6]. Recent work shows that human-AI collaborative workflows can produce personas that are more representative and empathy-evoking than those created by humans or AI alone [10]. Meanwhile, limitations of human participant research are well-documented: demand characteristics lead participants to align behavior with perceived researcher expectations [12, 13], and social desirability bias intensifies in face-to-face and high-status settings [1]. Despite this, the field still defaults to human feedback as ground truth. A less examined possibility is that AI personas, when grounded in research data, might sometimes deliver less biased insights than human participants in specific, identifiable contexts.

This paper presents a case from our practice as UX researchers at a Turkish fintech institution. After developing research-based AI personas, we conducted a workshop with actual users matching these persona profiles and obtained feedback that raised methodological concerns, prompting us to reconsider the assumed superiority of real human feedback.

We argue that the field should move beyond binary framings of AI versus human research toward understanding which biases each approach introduces, and when one might be preferable.

## 2 Developing Research-Based AI Personas

In 2024, our team interviewed roughly seventy banking customers, delving into their experiences. These customers represented different wealth levels and investment styles. Based on this research, we created six personas for our wealth management services.

Instead of keeping these documents static, we turned them into conversational AI agents using ChatGPT’s CustomGPT. Each agent was built using demographic information, financial goals, challenges, and behavioral patterns from our interview data.

The goal was to make these agents available throughout the company, allowing teams to access user perspectives without needing to schedule interviews. However, organizational concerns about AI credibility limited adoption.

## 3 Case Study: The Workshop on Design Thinking

We held a full-day design thinking workshop in Istanbul to talk about value propositions for a wealth management app that uses AI to give investment advice.

Forty high-net-worth clients took part, and their portfolio managers (PMs) helped find them. There were five tables with eight people at each one. The workshop was held at a fancy hotel, and the people who came stayed there. Importantly, PMs were there with their clients and took part in table discussions all morning. We thought about using our AI avatars with real people to get a better picture of the situation, but we decided to stick with only humans because we were worried about how stakeholders would react.

### 3.1 Observed Outcomes

After a C-level introduction that highlighted the strategic significance of wealth management innovation, participants undertook a structured design thinking process. This process included ice-breaking exercises that introduced the fundamentals of design thinking, persona presentations, and user journey mapping that explored investment behaviors. Subsequently, table representatives presented their findings to the entire group. Rather than directly asking about AI-augmented investment advice, we probed indirectly: What could happen if your PM was not available at that moment? and What happens if your PM leaves the company? Three of five tables produced answers indicating high dependency on existing PM relationships. Typical responses included I would follow him to wherever he goes and We are happy working with Mr. [name]. Questions about digital alternatives were consistently deflected toward human touch.

All three of these tables included participants with existing relationships to PMs who were present and actively participating in the discussions. In one case, a PM even served as the table’s representative during the playback session.

### 3.2 Identifying Sources of Bias

Analysis revealed multiple contaminating factors. Social desirability bias emerged because PMs were physically present at the tables; expressing criticism of human advisors or enthusiasm for AI alternatives would have been socially awkward when these managers had personally invited participants and were contributing to discussions. Context effects arose from the luxury setting, positioning participants as valued guests rather than critical evaluators. Relational anchoring occurred as participants were literally being asked to imagine losing someone sitting beside them. Acquiescence bias led toward agreement in the hospitable environment.

The uncomfortable insight: our AI personas were built from approximately 70 interviews conducted in neutral settings, including homes, coffee shops, and phone calls, without PMs present and without luxury accommodations signaling expected responses. Mapping each bias to the AI alternative reveals specific immunities: the personas could not engage in social desirability management because no authority figure was present in their interaction context; they were immune to luxury-setting context effects because they have no environmental perception; they could not relationally anchor because they hold no ongoing relationships with stakeholders; and they could not acquiesce to hospitality because they experience none. That research data, mediated through synthetic personas, may have contained less contextual contamination than what we obtained from carefully recruited human participants in our workshop.

## 4 AI Personas as Complement to Human Research

We do not claim AI personas are unbiased; they inherit biases from training data, grounding research, and configuration prompts [3, 4], and can exhibit demographic biases in age, occupation, and geographic representation [11]. Two limitations deserve direct comparison with the biases we observed. Sycophancy, where AI personas agree with or flatter the prompter, is structurally analogous to the social desirability bias our human participants exhibited. However, AI sycophancy is consistent and detectable through adversarial prompting, whereas social desirability in our workshop was situationally amplified by PM presence and luxury context in ways we could not anticipate or control. Hallucination, generating plausible but fabricated details, has no direct human analogue in our case, but introduces a risk of false specificity that must be weighed against the false consensus our workshop produced.

The critical distinction is not that AI personas are less biased overall, but that their biases are different and more predictable. An AI persona grounded in solid research cannot be influenced by stakeholder presence, does not feel pressure to be appreciative, cannot anchor to preserve relationships, and is unaffected by hospitality. These are precisely the biases that contaminated our workshop.

AI personas also offer continuous accessibility, available during sprint planning, design reviews, or any decision-making moment, potentially fostering more sustained engagement with user perspectives than periodic research permits.

### 4.1 A Preliminary Decision Heuristic

Our experience suggests the following preliminary guidance. AI personas may be preferable complements when: (a) stakeholders or gatekeepers will be present during research, creating social desirability pressure; (b) the research setting itself signals expected responses through luxury, hospitality, or corporate framing; (c) participants have ongoing relationships with organizational representatives that create relational anchoring; or (d) the research questions concern potentially face-threatening topics like replacing human services with AI. Conversely, human participants remain essential when: (a) emotional nuance and nonverbal cues are critical to understanding; (b) researchers seek genuinely unexpected insights beyond the training data; (c) behavioral observation rather than stated preference is the primary method; or (d) the research context itself is the object of study. We propose that teams use AI personas as a bias-resistant baseline, then triangulate with human research while explicitly accounting for contextual contamination.

## 5 Open Questions

Our experience surfaces several unresolved tensions we hope to explore with workshop participants. First, when is authentic authenticity? Our human participants were genuinely expressing their values; they likely do value their PM relationships. But these expressions were amplified by context while other perspectives were suppressed. If context always shapes expression, what does authentic feedback mean? Second, the validation paradox: to validate AI personas, we compare them against human responses, but if human responses are themselves contaminated by context, what serves as ground truth? Should we validate AI personas against behavior rather than stated preferences? Third, many organizations research their own customers in settings that introduce systematic biases, such as user conferences, VIP events, and advisory boards. These are often the only contexts where access is possible. Are AI personas a way to establish a bias-resistant baseline for these inherently compromised settings?

## 6 Conclusion

We entered our workshop expecting human participants would provide more authentic insights than AI personas. We emerged with a more complex understanding of authenticity in user research.

The participants were not dishonest; they likely did value their PM relationships. However, the setting amplified certain signals while suppressing others in unanticipated ways.

Our AI personas would not have been subject to those distortions. Different limitations, but not those ones.

We do not advocate replacing human research with AI personas. We believe the field needs guidance for recognizing when traditional research introduces biases that synthetic approaches might mitigate. Responsible use requires understanding their distinctive potential to complement human methods.

## ACKNOWLEDGMENTS

We thank colleagues in Digital Banking and Private Banking at Fibabanka for their collaboration.